\documentclass{article}

\usepackage{PRIMEarxiv}

\usepackage[utf8]{inputenc} 
\usepackage[T1]{fontenc}    
\usepackage{hyperref}       
\usepackage{url}            
\usepackage{booktabs}       
\usepackage{amsfonts}       
\usepackage{nicefrac}       
\usepackage{microtype}      
\usepackage{lipsum}
\usepackage{fancyhdr}       
\usepackage{graphicx}       
\graphicspath{{media/}}     

\title{Genetic Profile-Based Drug Sensitivity Prediction in Acute Myeloid leukemia Patients Using SVR
}

\author{
  Sadia Ruhama \\
  BRAC University \\
\texttt{sadia.ruhama@g.bracu.ac.bd}
}

\begin{document}
\maketitle

\begin{abstract}
Acute Myeloid Leukemia (AML) is a highly aggressive blood cancer, with low survival rates. Hence, emphasizing the importance of the urgent need for effective treatment modalities. In recent times, the advances in cancer genomics have increased our understanding of AML, as a result, enabling precision oncologists to develop personalized treatment based on individual genetic features and increase the survival rate. However, there is a lack of understanding how effectively genetic features can be used to predict which drugs are the most suitable for individual tailored treatment. Therefore, this study explores the potentiality of Support Vector Regression (SVR)  in predicting drug sensitivity of AML patients solely based on their genetic profile. The paper utilized a dataset from Genomics of Drug Sensitivity (GDSC) and developed a precise model that identified the most significant genetic features affecting drug response and achieved promising results with an R² score 0.9523 on the validation set and 0.8928 on the test set.
\end{abstract}


\section{Introduction}
AML is an aggressive form of blood cancer, with only 28.3 percent of adults surviving at least five years post-diagnosis, highlighting the urgent need for novel treatment strategies to improve patient outcomes \cite{lin2020ex}. Despite considerable research efforts, AML continues to pose significant challenges in both diagnosis and treatment. The emergence of genomics has recently enhanced our understanding of AML, offering insights that may lead to more effective approaches for managing this complex disease \cite{eckardt2020application}. Furthermore, AML has a higher incidence rate in the United States compared to the other three main leukemia subtypes acute lymphocytic leukemia (ALL), chronic myeloid leukemia (CML), and chronic lymphocytic leukemia (CLL), and accounts for the highest proportion of leukemia-related deaths \cite{shallis2019epidemiology}.

Machine learning has shown promise in analyzing high-dimensional biological data without requiring extensive expert input. In cancer prognosis, machine learning methods have yielded encouraging results, with various models, ranging from Artificial Neural Networks (ANNs) to Support Vector Machines (SVMs), successfully applied to support the development of personalized medicine \cite{10.1145/3290818.3290823}. Recent studies have demonstrated that Support Vector Regression (SVR) can predict optimal drug responses by analyzing the relationship between gene mutations and drug efficacy, thereby enhancing the precision of individualized treatment plans for cancer patients \cite{brindha2022precise}. Advances in genomic technologies have further enabled the development of personalized treatments by leveraging the genetic makeup of tumors, which can offer valuable insights into treatment response. In this context, analyzing gene co-expression networks has been shown to improve prediction accuracy \cite{ahmed2020network}.

Nevertheless, a major challenge in precision oncology lies in accurately predicting how tumors will respond to specific treatments \cite{baptista2021deep}. Research indicates that drug efficacy is closely tied to selecting the appropriate features, such as genetic mutations and drug targets \cite{koras2020feature}. Drug sensitivity prediction models not only facilitate personalized treatment strategies but also contribute to biomarker discovery and drug development. Despite advances in genomic data analysis, there remains a gap in fully understanding how effectively genomic profiles alone can guide the selection of the most suitable therapies for individual patients \cite{dawood2024cancer}.

This study investigates the potential of using only the genetic profiles of AML patients to predict drug sensitivity, with the goal of enabling personalized treatment. By employing SVR, the study aims to demonstrate the feasibility and effectiveness of drug response prediction based solely on genomic data.

\section{Literature Review}

AML is one of the most rapidly progressing cancers, affecting both pediatric and adult populations. It is defined by the overproduction of immature myeloid cells in the bone marrow and blood, which impairs normal hematopoiesis \cite{yamashita2020dysregulated}. Treatment of AML presents numerous challenges, including high costs and limited access to advanced diagnostic tools like flow cytometry and genetic testing due to financial constraints \cite{gomez2023acute}.



\subsection{Genetic Profiling in AML}
Genetic profiling plays a pivotal role in comprehending AML and its response to therapies. Identifying specific genetic characteristics can offer crucial insights into patient-tailored treatment strategies, thereby enhancing therapeutic outcomes \cite{tumkursattar2023investigate}. Several genetic alterations have been pinpointed as key contributors to AML pathogenesis and resistance to treatment. Mutations in genes such as FLT3, NPM1, and IDH1/2 have garnered significant attention owing to their implications for prognosis and therapeutic responses \cite{castano2025aml}. For instance, FLT3-ITD mutations are linked to poor prognosis and resistance to conventional chemotherapy, whereas NPM1 mutations often correlate with favorable treatment responses \cite{pan2023prognostic}. Conversely, TP53 mutations are associated with resistance to multiple targeted therapies and an elevated relapse risk \cite{wei2023characterizing}.

Numerous studies have investigated the utilization of genetic data to predict drug sensitivity in AML and other malignancies. A large-scale integration of genomic and drug response data from AML patients has demonstrated that specific genetic alterations can forecast responsiveness to targeted therapies \cite{tyner2018functional}. Similarly, analyses of genomic features across various cancer cell lines have established correlations between genetic mutations and drug efficacy, underscoring the potential of machine learning in precision oncology \cite{ren2024interpretable}.

\subsubsection{Support Vector Regression in Drug Sensitivity Prediction}
Support Vector Regression (SVR) is a robust machine learning technique particularly adept at predicting continuous variables, such as drug sensitivity metrics. SVR operates by mapping input data—such as genetic profiles—into a higher-dimensional space, where a hyperplane is constructed to optimally fit the data while minimizing prediction errors. This hyperplane is then employed to forecast continuous output values for novel data points. SVR's suitability for drug sensitivity prediction stems from the continuous nature of drug responses, which range from complete resistance to high sensitivity. Unlike classification models that yield discrete outcomes, SVR captures the nuanced, graded spectrum of drug efficacy across patient populations \cite{glanzer2024navigating}.

Recent advancements in 2024 and 2025 have further validated SVR's efficacy in oncology. For example, hybrid SVR models optimized with particle swarm algorithms have enhanced predictions in breast cancer \cite{brindha2022precise}. In AML-specific contexts, SVR has been integrated with network-driven frameworks to predict drug responses with high precision. Moreover, knowledge graph-guided SVR models have improved feature engineering by incorporating biological relationships, leading to more accurate predictions.

\subsection{Recent Advances in Machine Learning for AML}
The integration of multi-omics data with machine learning has seen significant progress. The BeatAML consortium, for instance, applied machine learning to fuse genomic, transcriptomic, and ex vivo drug sensitivity data from over 672 AML patients, resulting in robust predictive models. Deep learning approaches have also been explored for genomic-perspective drug response prediction, offering novel insights into mechanistic underpinnings.

In relapsed/refractory AML, machine learning strategies prioritize synergistic drug combinations, enhancing therapeutic efficacy. Network-based platforms like NetAML systematically develop drug-specific models, identifying sensitive AML subtypes. Furthermore, artificial intelligence has been leveraged to predict anti-cancer drug responses based on multi-omics data, addressing resistance mechanisms.

Emerging AI-driven precision therapies utilize systematic analyses of clinical and genomic data to forecast drug resistance phenotypes. Mutation patterns, particularly in genes like FLT3 and NPM1, have been shown to predict sensitivity to specific drug classes, such as MEK inhibitors. Ensemble-based models like MDREAM integrate omics data for comprehensive drug response predictions in AML.

These advancements underscore the transformative potential of machine learning in AML, supporting the adoption of SVR in this study for its balance of accuracy and computational efficiency.









\section{Methodology}
\subsection{Data Acquisition and Preprocessing}
The dataset used in this study was extracted from the Genomics of Drug Sensitivity in Cancer(GDSC). Particularly,  the data used here was obtained from GDSC2 dataset \cite{GDSC2} which provides insights into how cancer cell lines responds to a corresponding drug. Initially, this paper started with 952 sample where each sample representing unique observations of drug responses and genetic profile in AML where each row specifies information regarding how the drug interacts with genetic profile, including details such as drug name, a unique drug ID, the biological target which is important to understand the drugs mechanism and the target pathway that the drug influences is  listed. Additionally, features such as ic50 effect size were considered as target since it gives a valid measurement to evaluate the effectiveness of different drugs for treating AML. As the IC50 value demonstrates the amount of drug needed to stop half of the cancer cells from growing where it suggests a lower IC50 value simply indicates that the drug is more effective at fighting cancer \cite{berrouet2020comparison}. Finally, after all the data cleaning, the study was left with 902 samples from where the dataset were divided into 80 percent training and 20 percent testing. Following this, the second split consists of further dividing 80 percent of the data into training and validation, involving 64 percent of the original data for training and 16 percent for validation.

\subsection{Building the SVR model}

Feature selection constitutes a vital phase in model construction, particularly in high-dimensional genomic datasets. To identify the most influential features, Recursive Feature Elimination (RFE) was employed in conjunction with a linear SVR kernel. RFE iteratively trains the model, ranks features based on their coefficients, and eliminates the least significant until the desired subset is achieved. In this study, RFE was configured to select the top 10 features, focusing on those most predictive of drug responses.

The selected features were subsequently partitioned for training, validation, and testing. This dimensionality reduction not only alleviated computational burden but also emphasized relevant variables, enhancing model interpretability and performance.

Upon feature selection, the SVR model was refined through hyperparameter tuning using Grid Search combined with cross-validation. This exhaustive search explored a broad spectrum of values for the regularization parameter C, which balances training error minimization and model complexity, and the epsilon parameter, defining the margin of tolerance for errors without penalties. The radial basis function (RBF) kernel was adopted to capture linear relationships inherent in genomic data.

The optimization process identified the best parameter combination, yielding superior performance on the training data. Cross-validation was integrated to ensure robust and reliable model evaluation, mitigating overfitting risks. The mathematical formulation of SVR involves minimizing the objective function:

\[
\min_{w, b, \xi, \xi^*} \frac{1}{2} \|w\|^2 + C \sum_{i=1}^n (\xi_i + \xi_i^*)
\]

subject to:

\[
y_i - (w^T \phi(x_i) + b) \leq \epsilon + \xi_i, \quad (w^T \phi(x_i) + b) - y_i \leq \epsilon + \xi_i^*, \quad \xi_i, \xi_i^* \geq 0
\]

where \(\phi\) denotes the feature map induced by the RBF kernel: \(K(x_i, x_j) = \exp(-\gamma \|x_i - x_j\|^2)\).

\subsection{Evaluating Model Performance}
In order to validate the performance of the Support Vector Regression (SVR) model, this study utilized many different metrics such as Mean Squared Error (MSE) and R² scores on validation and test datasets, hence, providing insights into how well the model performs when it is predicting drug responses based on the selected genetic features. The Mean Squared Error (MSE) score on the validation set marked at 0.0066 and an R² of 0.9523 which indicates that 95.23 percent of the variance in the drug response was explained by the model. Moreover, on the test set the MSE was 0.0096, and the R² score was 0.8928, indicating an accuracy of about 89.28 percent.The cross-validation scores ranged from 0.9386 to 0.9876, with a mean cross-validation score of 0.9642, suggesting that the model performs consistently well across different subsets of the data.

\section{Findings}
Through Recursive Feature Elimination (RFE), the SVR model identified the 10 most significant features, which improved the model's performance and also helped determine which specific genetic profiles of AML patients were most predictive of their response to treatment. The model showcases promising results by achieving R² score of 95.23 percent on the validation set and 89.28 percent on the test set. These are shown briefly in Table 2 and Table 3.

\begin{figure}[h]
    \centering
    \includegraphics[width=0.5\textwidth]{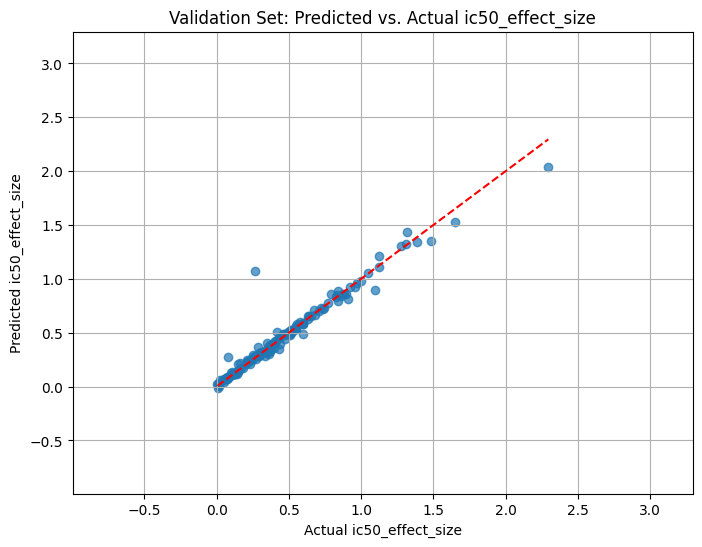} 
    \caption{Validation set}
    \label{fig:example_image}
\end{figure}

\begin{figure}[h]
    \centering
    \includegraphics[width=0.5\textwidth]{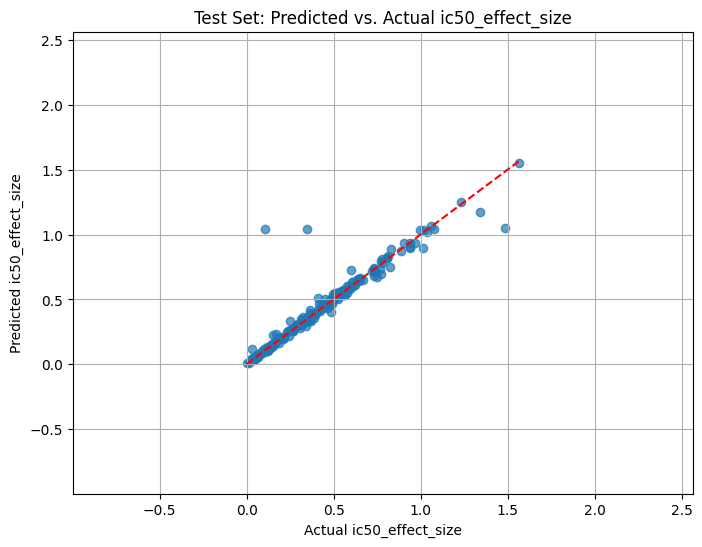} 
    \caption{Test set}
    \label{fig:example_image}
\end{figure}

The scatter plot as shown in figure 1 and figure 2 illustrates the relationship between predicted vs. Actual IC50 values in validation and test set. The x-axis represents the actual IC50 values, while the y-axis displays the predicted IC50 values for the samples. IC50 values are a measurement used to predict how effectively a drug can inhibit cancer growth by telling us the concentration of a drug needed to reduce the growth of cancer cells by 50 percent, also a lower IC50 values indicates the drug is effective \cite{joo2019deep}.  The validation set suggests a strong correlation as shown in the graph where the points are observed to cluster to the diagonal line. Also in the test set the points are clustered together.

\begin{table}[h]
    \centering
    \begin{tabular}{|c|c|}
        \hline
        \textbf{Parameter} & \textbf{Value} \\
        \hline
        C & 100 \\
        $\epsilon$ & $0.01$ \\
        Kernel & rbf \\
        \hline
    \end{tabular}
    \caption{Best Parameters Identified for the Model}
\end{table}

\begin{table}[h]
    \centering
    \begin{tabular}{@{}ll@{}}
        \toprule
        \textbf{Metric} & \textbf{Value} \\ \midrule
        Best cross-validation score & 0.9659 \\
        Validation Mean Squared Error (MSE) & 0.0066 \\
        Validation R\textsuperscript{2} Score & 0.9523 \\
        Test Mean Squared Error (MSE) & 0.0096 \\
        Test R\textsuperscript{2} Score & 0.8928 \\ \bottomrule
    \end{tabular}
    \caption{Model performance metrics}
    \label{tab:performance_metrics}
\end{table}

\begin{table}[h]
    \centering
    \begin{tabular}{@{}ll@{}}
        \toprule
        \textbf{Cross-Validation Scores} &  \\ \midrule
        1 & 0.9612 \\
        2 & 0.9876 \\
        3 & 0.9754 \\
        4 & 0.9386 \\
        5 & 0.9582 \\ \midrule
        \textbf{Mean Score} & 0.9642 \\ \bottomrule
    \end{tabular}
    \caption{Cross-validation scores}
    \label{tab:cross_validation_scores}
\end{table}

Furthermore, the features mentioned in this paper have the potential to be targeted as biomarkers, such as SACS mut, THR-103, and glutathione. So the presence or absence of these drug(THR-103, glutathione) might correlate with specific outcomes. 

\section{Discussion}

The SVR model demonstrated strong predictive capabilities, achieving an R² score of 0.9523 on the validation set and 0.8928 on the test set. This indicates that genetic profiles can be effectively used to predict drug sensitivity in AML patients. The study confirms that IC50 values can be effectively predicted using machine learning, reinforcing previous findings that genetic mutations play a significant role in personalized cancer treatment. Removing the features KRAS, NRAS, and ASXL1 from the dataset did not alter the Mean Squared Error (MSE), which remained at 0.0123. This suggests that these particular features may not significantly influence the model's predictive capability. When comparing the SVR model to a Random Forest Regressor, the latter produced a slightly higher MSE of 0.0128, indicating that SVR with an RBF kernel performs better for this dataset. Additionally, the impact of data scaling was evident when an SVR model trained without feature scaling resulted in a substantial increase in MSE to 0.1006. This highlights the importance of scaling input data, as unscaled data can significantly reduce model accuracy.

Further analysis of kernel functions within the SVR framework revealed that the polynomial kernel had an MSE of 0.0185. RBF kernel is better suited for capturing these complexities. The polynomial kernel, while accounting for some nonlinearity, did not match the performance of the RBF kernel. These findings emphasize the critical role of kernel selection, feature inclusion, and data preprocessing in developing accurate SVR models.

Furthermore, studies have shown that the integration of omics data and machine learning approaches, including SVM-based models, can improve drug response predictions \cite{Trac2023}. Similarly, another study showcases that machine learning models using RNA sequencing and clinical data can predict drug response in AML patients, as a result, improving personalized treatment \cite{karathanasis2024machine}. Many researchers have used machine learning to predict how well drugs work for AML patients, focusing on specific gene mutations like FLT3, NPM1, and IDH1/2, for example, a study looked at how FLT3 mutations affect AML patients who also have NPM1 mutations \cite{Chan2022FLT3}. They found that one type of FLT3 mutation (FLT3-ITD) makes the disease worse, while another type (FLT3-TKD) is less understood. This shows that it is important to look at the exact type of mutation when predicting drug responses. Another study explored how machine learning can help in AML treatment \cite{Eckardt2020MLAML}. They found that using AI with genetic information makes it easier to predict which drugs will work best for each patient, leading to more personalized treatments. These studies support the idea that AI and genetic data together can help doctors make better treatment decisions for AML patients.

This study utilizes the GDSC2 dataset, a well-established and publicly available resource, ensuring that the findings are reliable and can be reproduced by other researchers. The use of such a high-quality dataset strengthens the credibility of the results and allows future studies to build upon this work. To improve efficiency, the study employs Recursive Feature Elimination (RFE) for feature selection. This method helps identify the most relevant genetic features while reducing unnecessary data, making the model both more efficient and less computationally demanding. By optimizing the number of features without compromising accuracy, the study ensures that the model remains practical for real-world applications. Another key strength of this study is its use of Support Vector Regression (SVR) instead of traditional classification models. Many existing approaches categorize drug response as either effective or ineffective, but in reality, drug sensitivity is a continuous measure. By predicting drug response values on a spectrum rather than as binary categories, this study provides a more nuanced and clinically relevant model that better reflects real-world patient responses.

\subsection{Limitations}
Despite its strengths, the study has some limitations. One of the main challenges is the dataset size. Although the GDSC2 dataset is reliable, it contains only 902 AML-specific samples. A larger and more diverse dataset could improve the model’s ability to generalize across different patient populations and enhance its robustness.

Additionally, the model is based solely on genetic data, but drug sensitivity is influenced by other factors, such as epigenetics, interactions within the tumor microenvironment, and individual patient physiology. Since these aspects were not considered, the model may not fully capture all determinants of drug response. Incorporating additional biological factors in future studies could lead to more comprehensive predictions.

Another challenge is the interpretability of the model. While SVR performs well in predicting drug response, it does not inherently explain how specific genetic features contribute to these predictions. This limits its usefulness in clinical decision-making. Techniques like SHAP (Shapley Additive Explanations) or LIME (Local Interpretable Model-agnostic Explanations) could be integrated in future studies to provide more transparency and improve understanding of how different genetic factors influence drug response.

\section{Conclusion and Future Work}
In conclusion, this paper provides a detailed investigation in predicting drug sensitivity of AML patients using SVR model that focuses solely on genetic profile. Moreover, the key findings of the study indicates the potential of machine learning, particular SVR model in revolutionizing cancer treatment by enabling tailored treatment. Besides, the ability of the model to understand and find patterns between the non linear relationship of genetic profile and drug sensitivity provides new insights into how treatments can be optimized for individual patients, holding promising results for precision medicine in oncology, particularly for AML, where effective treatment is often hindered due to limitations in genetic variability.

The SVR model achieved high accuracy suggesting that only genetic profile can predict drug sensitivity in AML patients, as a result, enabling oncologists to tailor personalized treatment. Into the bargain, the study showcases the options in incorporating genetic profile and multi-omic data, such as, proteomics or transcriptomics, hence, opening the door for future research. Aside from that, the study also encourages to apply this technique in other aspects of cancer and see whether different cancer subtype require different adjustments, considering each cancer type has unique genomic features.


\bibliographystyle{unsrt}  
\bibliography{references}

\end{document}